\documentstyle[12pt]{article}
\begin{document}
\newcommand{\beq}{\begin{equation}}

\newcommand{\eeq}{\end{equation}}
\newcommand{\beqa}{\begin{eqnarray}}
\newcommand{\eeqa}{\end{eqnarray}}
\newcommand{\bea}{\begin{eqnarray}}
\newcommand{\eea}{\end{eqnarray}}
\newcommand{\no}{\nonumber}
\newcommand{\grts}{\greaterthansquiggle}
\newcommand{\lets}{\lessthansquiggle}
\newcommand{\ul}{\underline}
\newcommand{\ol}{\overline}
\newcommand{\ra}{\rightarrow}
\newcommand{\Ra}{\Rightarrow}
\newcommand{\ve}{\varepsilon}
\newcommand{\vp}{\varphi}
\newcommand{\vt}{\vartheta}
\newcommand{\dg}{\dagger}
\newcommand{\wt}{\widetilde}
\newcommand{\wh}{\widehat}
\newcommand{\dfrac}{\displaystyle \frac}
\newcommand{\fsl}{\not\!}
\newcommand{\ben}{\begin{enumerate}}
\newcommand{\een}{\end{enumerate}}
\newcommand{\bfl}{\begin{flushleft}}
\newcommand{\efl}{\end{flushleft}}
\newcommand{\ba}{\begin{array}}
\newcommand{\ea}{\end{array}}
\newcommand{\btab}{\begin{tabular}}
\newcommand{\etab}{\end{tabular}}
\newcommand{\bit}{\begin{itemize}}
\newcommand{\eit}{\end{itemize}}
\newcommand{\be}{\begin{equation}}
\newcommand{\ee}{\end{equation}}
\newcommand{\bearr}{\begin{eqnarray}}
\newcommand{\eearr}{\end{eqnarray}}
\newcommand{\per}{\;\;.}
\newcommand{\pl}{PL}
\newcommand{\spi}{s_\pi}
\newcommand{\ql}{QL}
\newcommand{\qn}{QN}
\newcommand{\pn}{PN}
\newcommand{\qq}{Q^2}
\newcommand{\mee}{m_l^2}
\newcommand{\thp}{\theta_\pi}
\newcommand{\mtiny}[1]{{\mbox{\tiny #1}}}
\newcommand{\MS}{\mtiny{MS}}
\newcommand{\GeV}{\mbox{GeV}}
\newcommand{\MeV}{\mbox{MeV}}
\newcommand{\keV}{\mbox{keV}}
\newcommand{\ren}{\mtiny{ren}}
\newcommand{\kin}{\mtiny{kin}}
\newcommand{\hint}{\mtiny{int}}
\newcommand{\tot}{\mtiny{tot}}
\newcommand{\CHPT}{\mtiny{CHPT}}
\newcommand{\QED}{\mtiny{QED}}
\newcommand{\syst}{\mbox{syst.}}
\newcommand{\stat}{\mbox{stat.}}
\newcommand{\wave}{\mbox{wave}}
\newcommand{\co}{\; \; ,}
\newcommand{\nn}{\nonumber \\}
\newcommand{\fff}{\bar{f}}
\newcommand{\ffg}{\bar{g}}

\renewcommand{\theequation}{\arabic{section}.\arabic{equation}}
\renewcommand{\thetable}{\arabic{table}}

\begin{titlepage}

\def\mytoday#1{{ } \ifcase\month \or
 January\or February\or March\or April\or May\or June\or
 July\or August\or September\or October\or November\or December\fi
 \space \number\year}

\begin{flushright}
BUTP--94/4

ROM2F 94/05

hep-ph/9403390
\end{flushright}

\vspace*{1cm}
\begin{center} {\Large \bf $K_{l 4}$ - DECAYS BEYOND ONE LOOP}$^\sharp$

\vspace{3cm}
{\bf{J. Bijnens}$^a$, \bf{G. Colangelo}$^{b,c}$} and {\bf{J. Gasser}$^b$}

\vspace{2cm}

\mytoday \\
\vfill
\end{center}

\begin{abstract}

\noindent
The matrix elements for $K\rightarrow \pi \pi \l \nu$ decays are described by
 four form factors $F,G,H$ and $R$. We complete previous calculations by
evaluating $R$ at next-to-leading order in the low-energy expansion.
We then estimate higher order contributions  using dispersion
relations and determine the low-energy constants $L_1,L_2$ and $L_3$ from
data on $K_{e4}$ decays and on elastic pion scattering. Finally, we present
 predictions for the slope of the form factor $G$ and for total decay
rates.

\noindent

{\underline{\hspace{5cm}}}\\

\noindent
$\sharp$ Work supported in part by Schweizerischer Nationalfonds/ Bundesamt
f\"ur Bildung und Wissenschaft (BBW)/
 EEC Human Capital and Mobility
Program/
 Fondazione
"Angelo della Riccia"/ INFN.

\noindent
a)
NORDITA,
 Blegdamsvej 17,
DK-2100 Copenhagen, Denmark.

\noindent
b)
Universit\"at Bern, Sidlerstrasse 5,
CH$-$3012 Bern, Switzerland.

\noindent
c)
Dipartimento di Fisica,
Universit\`{a} di Roma II - "Tor Vergata",
Via della

\noindent
\hspace{3.1mm} Ricerca Scientifica 1, I-00173 Roma, Italy.

\vspace{.15cm}

\end{abstract}
\end{titlepage}
\vfill \eject

\pagestyle{plain}
\clearpage
\setcounter{page}{2}
\setcounter{equation}{0}
\setcounter{subsection}{0}
\setcounter{table}{0}
\setcounter{figure}{0}
\section{Introduction}
\label{Intro}

In this article we analyze $K_{l4}$ decays,
\be
K\rightarrow \pi\pi l \nu \; ; \;l=e,\mu \co
\label{in1}
\ee
in the framework of chiral perturbation theory (CHPT). This method, also called
"energy expansion" in the following, is based on  an expansion of the Green
functions  in powers of the external momenta and of the light quark masses
 \cite{wein79}-\cite{revchir}. The matrix elements for $K_{l4}$ processes are
described by four  form factors $F,G,H$ and $R$. Their energy expansion reads
\be
I=\frac{M_K}{F_\pi}\left\{I^{(0)} + I^{(2)} + I^{(4)} +\cdots \right\}\;;\;
I=F,G,H,R\co
\label{in2}
\ee
where $I^{(n)}$ is a quantity of order $E^n$.
The predictions for the lowest
order terms  were first
given by Weinberg \cite{weinkl4}.
The anomaly contribution $H^{(2)}$ has been determined in \cite{wesszkl4},
whereas $F^{(2)}$ and $G^{(2)}$  have been evaluated in
 \cite{bijnenskl4,riggen}. For a calculation of $H^{(4)}$ see \cite{llamet}.
The experimental results \cite{ross} for $F,G$  turn out to be $30-50\%$  above
 the leading contributions.
The missing piece must then come from higher orders.
 The expressions for $F^{(2)},G^{(2)}$  involve  the low-energy
constants
$L_1,\ldots,L_9$, which are not fixed by chiral symmetry alone and which
must be determined phenomenologically at the present stage of our capability
to solve QCD. This was done in \cite{glnpb}, where the $L_i$'s were
pinned down using experimental data (not related to $K_{l4}$ decays) and
involving the large-$N_C$ prediction which states that, in the limit where
the number of colors becomes large, certain combinations of the low-energy
constants are suppressed.
The decay (\ref{in1}) is the simplest process where this  rule can be
tested \cite{bijnenskl4,riggen}. In addition, it allows one to perform an
independent determination of  $L_1,L_2$ and $L_3$ and thus to
check consistency with other data.

                                                        The aim of the present
article is threefold. First, we fill the gap in the literature and evaluate
also
the next-to-leading order term $R^{(2)}$.
 [The amplitude R  is completely negligible in $K_{e4}$ decays, because its
contribution to the rate is suppressed by the factor $m_l^2$. It  must be
retained, however, in
the $K_{\mu 4}$ channel ]. Second, we note that, because the strange quark mass
is not very small on a typical hadronic scale, the corrections $I^{(2)}$ to the
leading-order terms of the form factors are substantial. The determination of
the $L_i$'s from $K_{l4}$ decays is therefore affected with  substantial
uncertainties
if carried out using only the first two terms in the expansion (\ref{in2}),
 as was done in
\cite{bijnenskl4,riggen}. Here we improve these calculations by estimating the
size of higher order contributions to $F$. We use
for this purpose the method developed
in \cite{dgl}, which amounts to write a dispersive representation for the
relevant
partial wave amplitudes, fixing the corresponding subtraction constants with
chiral perturbation theory. We are then able to reduce the
uncertainties in the determination of $L_1,L_2$ and $L_3$ in a significant
manner, even more  so if data on elastic $\pi\pi$
scattering is considered in addition.
Third, we  predict the slope of the form
factor $G$, and show that we  may evaluate total decay rates for
all channels in $K_{l4}$ decays within rather small uncertainties, provided the
leading $S$- and $P$- wave form factors have  been determined
 experimentally
e.g. from $K^+\rightarrow \pi^+\pi^- e^+\nu_e$ decays.

 The plan of the paper is as follows. In section 2, we provide the
necessary kinematics and the definition of the form factors. Section 3 contains
the result of the one-loop calculation of these quantities.
In section 4, we use dispersion relations to construct a $I=0 \;S$-wave
amplitude
which has the correct phase to higher orders in the low-energy expansion. In
section 5, we use this improved amplitude to determine the low-energy constants
$L_1,L_2$ and $L_3$.
 Section
6 contains  predictions, whereas a summary and concluding remarks are
presented in section 7.
\setcounter{equation}{0}
\setcounter{subsection}{0}
\setcounter{table}{0}
\setcounter{figure}{0}

\section{Kinematics and form factors}

\label{kin}

We discuss the decays
         \bearr
          K^+(p) & \rightarrow & \pi^+(p_1) \;\pi^-(p_2) \;l^+(p_l) \; \nu_l
(p_{\nu})\co          \label{k1} \\
          K^+(p) & \rightarrow & \pi^0(p_1)\; \pi^0(p_2) \; l^+(p_l) \;\nu_l
(p_{\nu})  \co       \label{k2}\\
          K^0(p) & \rightarrow & \pi^0(p_1)\; \pi^-(p_2)\; l^+(p_l) \;\nu_l
(p_{\nu})         \label{k3} \;\; ; \; \;l=e,\mu \co
        \eearr
        and their charge conjugate modes.
 We do
        not consider isospin violating contributions and
        correspondingly set $m_u = m_d$, $\alpha_\QED = 0$.

        \subsection{Kinematics}

        We start with the process (\ref{k1}).
       The full kinematics of this decay requires five variables.
       We will use the
       ones introduced by Cabibbo and Maksymowicz \cite{cabmak}. It is
       convenient to
       consider three reference frames, namely the $K^+$ rest system
       $(\Sigma_K)$, the
       $\pi^+
       \pi^-$ center-of-mass system $(\Sigma_{2 \pi})$ and the
       $l^+\nu_l$ center-of-mass system $(\Sigma_{l \nu})$. Then
       the variables are

       \begin{enumerate}
       \item $s_\pi$, the effective mass squared of the dipion system,

       \item $s_l$, the effective mass squared of the dilepton system,

       \item $\theta_\pi$, the angle of the $\pi^+$ in $\Sigma_{2\pi}$
       with respect to the dipion line of flight in $\Sigma_K$,

       \item $\theta_l$, the angle of the $l^+$ in $\Sigma_{l\nu}$
       with  respect to the dilepton line of flight in
       $\Sigma_K$, and

       \item $\phi$, the angle between the plane formed by the pions
       in
       $\Sigma_K$ and the corresponding plane formed by the dileptons.

       \end{enumerate}

       The angles $\theta_\pi$, $\theta_l$ and $\phi$ are displayed in
       Fig. 1.

      The range of the variables is
      \bearr
      4 M^2_\pi & \leq & s_\pi = (p_1+p_2)^2 \leq (M_K - m_l)^2 \co
      \nonumber \\
      m^2_l & \leq & s_l =(p_l+p_\nu)^2 \leq (M_K - \sqrt{s_\pi})^2\co
      \nonumber \\
      0 & \leq & \theta_\pi, \theta_l \leq \pi, 0 \leq \phi \leq 2
      \pi. \label{h6}
      \eearr

It is useful to furthermore introduce the following combinations of four
vectors
\be
P=p_1+p_2, \; \; Q=p_1-p_2,\; \; L=p_l+p_\nu, \; \; N=p_l-p_\nu .
\ee
Below we will also use the variables
\be
t=(p_1-p)^2 ,u=(p_2-p)^2 ,\nu=t-u.
\ee
These are related to $s_\pi,s_l$ and $\theta_\pi$ by
\bearr
t+u&=&2M_\pi^2 +M_K^2 +s_l -s_\pi \; , \nonumber \\
\nu&=& -2\sigma_\pi X \cos\theta_\pi \; ,
\eearr
where
\bearr
\sigma_\pi &=& (1-4M_\pi^2/s_\pi)^\frac{1}{2}\, ,  \nonumber \\
            X&=&\frac{1}{2}\lambda^{1/2}(M_K^2,s_\pi,s_l)\co \nn
\lambda (x,y,z) &=& x^2+y^2+z^2 -2(x y +x z +y z)\per
\eearr
In addition we define
\be
\Sigma = M_K^2 +M_\pi^2.
\ee

      \subsection{Matrix elements and decay rates}

      The matrix element for $K^+ \rightarrow \pi^+ \pi^- l^+ \nu_l$
      is
      \be
      T = \frac{G_F}{\sqrt{2}} V^\star_{us} \bar{u} (p_\nu) \gamma_\mu
      (1-\gamma_5) \nu(p_l) (V^\mu - A^\mu) \co
      \label{k11}
      \ee
      where\footnote{
 In order to agree with the notation used by Pais and
Treiman \cite{paistr} and by Rosselet et al. \cite{ross}, we have changed the
previous  convention \cite{bijnenskl4,riggen} in the definition of the anomaly
form factor $H$. }

      \bearr
      I_\mu & = & < \pi^+ (p_1) \pi^- (p_2) \mbox{out}\mid
      I_\mu^{4-i5} (0) \mid K^+ (p)  >;\; I = V,A \co
      \nonumber \\
      V_\mu & = & - \frac{H}{M^3_K} \epsilon_{\mu \nu \rho \sigma} L^\nu
      P^\rho Q^\sigma \co
      \nonumber \\
      A_\mu & = & -i\frac{1}{M_K} \left [ P_\mu F +
      Q_\mu G + L_\mu R \right ] \co
      \label{k12}
      \eearr
and $\epsilon_{0123}=1$.
 The matrix elements for the other channels
(\ref{k2}, \ref{k3}) may be obtained from (\ref{k11}, \ref{k12}) by isospin
symmetry, see below.

      The form factors $F,G,R$ and $H$ are
analytic functions of the variables $s_\pi, t$ and $u$.
      The partial decay rate for (\ref{k1}) is given by

      \be
      d\Gamma = \frac{1}{2M_K(2\pi)^8} \sum_{spins} \mid
      T \mid^2 \delta^4(p-P-L) \frac{d^3p_1}{2p_1^0}  \frac{d^3p_2}{2p_2^0}
\frac{d^3p_l}{2p_l^0} \frac{d^3p_\nu}{2p_\nu^0} .      \label{k13}
      \ee

The quantity $\sum_{spins} \mid T \mid ^2$ is a Lorentz invariant quadratic
form in $F,G,R$ and $H$. All scalar products can be expressed
 in terms of the 5 independent variables $s_\pi,s_l,\theta_\pi,\theta_l$ and
$\phi$,  such that
\be
\sum_{spins} \mid T \mid ^2 = {2G_F^2 \mid V_{us} \mid ^2}{M_K^{-2}}
J_5(s_\pi,s_\l,\theta_\pi,\theta_l,\phi) \; \; .
\label{k13a}
\ee
Carrying out the integrations over the remaining $4 \cdot 3 - 5 = 7$ variables
in (\ref{k13})       gives \cite{cabmak}
 \bearr\label{k14}
      d \Gamma_5 &=& G^2_F \mid V_{us} \mid^2  N(s_\pi, s_l) J_5
      (s_\pi, s_l, \theta_\pi, \theta_l, \phi) ds_\pi ds_l d (\cos
      \theta_\pi) d(\cos \theta_l) d\phi \co \nn
      N(s_\pi, s_l) &=& (1-z_l) \sigma_\pi X
      /(2^{13} \pi^6 M_K^5) \co
      \eearr
      where
\bearr
J_5 &=&2(1-z_l)\left[ I_1 + I_2 \cos 2 \theta_l + I_3 \sin^2 \theta_l
 \cdot \cos 2
\phi + I_4 \sin 2 \theta_l \cdot \cos \phi \right. \nonumber \\
&+& \left. I_5 \sin \theta_l \cdot
\cos \phi
+  I_6 \cos \theta_l + I_7 \sin \theta_l \cdot \sin \phi + I_8 \sin 2
\theta_l \cdot \sin \phi \right. \nonumber \\
 &+& \left. I_9 \sin^2 \theta_l \cdot \sin 2 \phi \right] \co
\nonumber
      \eearr
with
      \bearr
I_1 &=& \frac{1}{4}\left\{ (1 + z_l) |F_1|^2  + \frac{1}{2}
(3+z_l)\left(|F_2|^2 +
|F_3|^2 \right) \sin^2 \thp   + 2z_l  |F_4|^2 \right\}
\co \nn
I_2 &=& - \frac{1}{4} (1-z_l)\left\{ |F_1|^2 - \frac{1}{2}
\left( |F_2|^2 + |F_3|^2 \right)  \sin^2 \thp  \right\}
\co \nn
I_3 &=& - \frac{1}{4} (1-z_l) \left\{ |F_2|^2 - |F_3|^2 \right\}\sin^2 \thp
\co \nn
 I_4 &=& \frac{1}{2}(1-z_l)\mbox{ Re} (F_1^* F_2)  \sin \thp
\co \nn
I_5 &=& -\left\{ \mbox{ Re} (F_1^* F_3) +
z_l  \mbox{ Re} (F_4^* F_2) \right\}  \sin \thp
\co \nn
I_6 &=&
- \left\{\mbox{ Re} (F_2^* F_3)\sin^2 \thp - z_l \mbox{ Re} (F_1^* F_4)\right\}
\co \nn
I_7 &=& - \left\{ \mbox{ Im} (F_1^* F_2) +
z_l  \mbox{ Im} (F_4^* F_3) \right\} \sin \thp
\co \nn
I_8 &=& \frac{1}{2} (1-z_l) \mbox{ Im} (F_1^* F_3) \sin \thp
\co \nn
I_9 &=& -\frac{1}{2}(1-z_l) \mbox{ Im} (F_2^* F_3) \sin^2 \thp \co
      \eearr
and
      \bearr\label{ki30}
 F_1& =& X \cdot F + \sigma_\pi (P L) \cos \thp \cdot G
\co \nn
 F_2& =& \sigma_\pi \left( s_\pi s_l \right)^{1/2} G
\co \nn
 F_3& =&  \sigma_\pi X \left( s_\pi s_l \right)^{1/2}  \frac{H}{M_K^2}
\co \nn
 F_4& =&- (P L)  F - s_l R - \sigma_\pi X \cos \thp \cdot G \; \; .
      \eearr
      The definition of $F_1, \ldots, F_4$  corresponds
      to the combinations used by Pais and Treiman \cite{paistr} (the
      different sign in the terms $\sim PL$
 is due to our use of the  metric
      $\mbox{diag} (+ ---)$). The form factors $I_1, \ldots, I_9$
      agree with the expressions given in \cite{paistr}.

      \subsection{Isospin decomposition}

      The $K_{l4}$ decays (\ref{k2}) and (\ref{k3}) involve the same
      form factors as displayed in Eq. (\ref{k12}). We denote by
      $A_{+-}$, $A_{00}$ and $A_{0-}$ the current matrix elements of
      the processes (\ref{k1})-(\ref{k3}). They are related by
      isospin symmetry\footnote{We use the  Condon-Shortley phase
conventions. Notice that we evaluate matrix elements and decay rates
for $K^0$ -- they differ from the corresponding $K_L$-quantities by a
normalization factor.},       \be
      A_{+-} =\frac{A_{0-}}{\sqrt{2}} - A_{00} \; \; .
      \label{i1}
      \ee

      This relation also holds for the individual form factors, which
      may be decomposed into a symmetric and an antisymmetric part under
      $t \leftrightarrow u \; (p_1 \leftrightarrow p_2)$. Because of Bose
      symmetry and of the $\Delta I = \frac{1}{2}$ rule of the
      relevant  weak currents, one has
      \bearr
      (F,G,R,H)_{00} & = & - (F^+, G^-, R^+, H^-)_{+-} \co
      \nonumber \\
      (F,G,R,H)_{0-} & = & \sqrt{2} (F^-, G^+, R^-, H^
      +)_{+-} \co
      \label{i2}
      \eearr
      where
      \be
      F^\pm_{+-} = \frac{1}{2} [F(s_\pi, t, u) \pm F(s_\pi, u, t)] \co
      \label{i3}
      \ee
      and $F(s_\pi, t, u)$ is defined in Eq. (\ref{k12}). $G^\pm,R^\pm$ and
$H^\pm$ are defined similarly.
The isospin relation for the decay rates is

\be
\Gamma (K^+\rightarrow \pi^+\pi^-l^+\nu_l) = \frac{1}{2}
\Gamma (K^0\rightarrow \pi^0\pi^-l^+\nu_l) +2  \Gamma (K^+\rightarrow
\pi^0\pi^0l^+\nu_l) \per
\ee

      \subsection{Partial wave expansion}

      The form factors may be written in a partial wave expansion in
      the variable $\theta_\pi$.
      We consider a definite isospin $\pi \pi$ state. Suppressing isospin
indices, one has \cite{berends}
      \bearr
      F &= & \sum^{\infty}_{l=0} P_l (\cos \theta_\pi) f_l - \frac{
      \sigma_\pi PL}{X} \cos
      \theta_\pi G \co
      \nonumber\\
      G & = & \sum^{\infty}_{l=1} P_l' (\cos \theta_\pi) g_l \co
      \nonumber \\
      R & = & \sum^{\infty}_{l=0} P_l (\cos \theta_\pi) r_l +
      \frac{ \sigma_\pi s_\pi}{X}
      \cos \theta_\pi G \co
      \nonumber \\
      H & = & \sum^{\infty}_{l= 0} P_l' (\cos \theta_\pi) h_l \co
      \label{i4}
      \eearr
      where
      \be
      P_l'(z) = \frac{d}{dz} P_l(z) \; \; .
      \label{i10}
      \ee

      The partial wave amplitudes $f_l, g_l, r_l$ and $h_l$ depend on
      $s_\pi$ and $s_l$. Their phase coincides with the phase shifts
      $\delta^I_l$ in elastic $\pi \pi$ scattering (angular momentum
      $l$, isospin $I$).
      More precisely, the quantities
      \bearr
     && e^{-i \delta^0_{2l}} X_{2l} \co
      \nonumber \\
     && e^{-i\delta^1_{2l+1}} X_{2l+1}\; ; \;l=0,1, \ldots \; ;
      \; X = f,g,r,h \co
      \label{i11}
      \eearr
      are real in the physical region of $K_{l4}$ decay (in our overall phase
      convention). The form
      factors $F_1$ and $F_4$ therefore have a simple expansion,
      \bearr
      F_1 &=& X \sum_{l} P_l
      ( \cos \theta_\pi) f_l \co
      \nonumber \\
      F_4& =& - \sum_{l} P_l (\cos \theta_\pi) (PL f_l + s_l r_l).
      \label{i12}
      \eearr

\renewcommand{\theequation}{\arabic{section}.\arabic{equation}}
\renewcommand{\thetable}{\arabic{table}}

\setcounter{equation}{0}
\setcounter{subsection}{0}
\setcounter{section}{2}

      \section{Theory}

      The theoretical predictions of $K_{l4}$ form factors have a long
      history which started in the sixties with the current algebra
      evaluation of $F$, $G$, $R$ and $H$. For an early review of the
      subject and for references to work prior to CHPT we refer the
      reader to \cite{chounet} (see also \cite{shabkl4}).
      Here we concentrate
on the evaluation of the form factors in the framework of CHPT
      \cite{bijnenskl4,riggen,stern}.

\vspace{.5cm}

     {\subsection{Tree level}}

\vspace{.5cm}
The chiral representation of the form factors at leading order  was
originally given by Weinberg \cite{weinkl4},
\bearr
F&=&G=\frac{M_K}{\sqrt{2}F_\pi}=3.74 \co \nonumber \\
R&=& \frac{M_K}{2\sqrt{2}F_\pi}\left(\frac{s_\pi+\nu}{s_l-M_K^2} +1
\right) \co \nonumber \\ H&=&0 \per
\label{t1}
\eearr
 The next-to-leading order corrections  are displayed
 below, and the later sections contain an estimate of yet higher order
contributions. Here we note that the
 total decay rates which follow from Eq. (\ref{t1}) are typically
a factor of two
(ore  more) below the data. As an example, consider the channel $K^+\rightarrow
\pi^+\pi^-e^+\nu_e$. Using (\ref{t1}), the total decay rate
becomes\footnote{If not stated otherwise, we use $F_\pi=93.2$ MeV,
$|V_{us}|=0.22$ and
$(M_\pi,M_K)=(139.6,493.6)$ MeV, $(135,493.6)$ MeV and $(137,497.7)$ Mev for
the decays (\ref{k1}), (\ref{k2}) and (\ref{k3}), respectively.} 1297
sec$^{-1}$,
whereas the experimental value is 3160$\pm$140 sec$^{-1}$ \cite{pdg}.

\subsection{{The form factors at one-loop}}

      In Ref. \cite{bijnenskl4,riggen}, the form factors $F$, $G$ and
      $H$ have been evaluated in CHPT at order $p^4$ (see also \cite{beg}).
We complement these works
with the evaluation of $R$ at the same order. Below we display the result of
our calculation, referring the reader to the above references and to
available reviews \cite{revchir} for the details of the methods used.

In order to make this article reasonably self-contained, we display the result
of all four form factors. The result for $F$ may be written in the form
      \be
      F(s_\pi, t, u)  =  \frac{M_K}{\sqrt{2} F_\pi} \left\{ 1 +
      \frac{1}{F_\pi^2}(U_F +P_F +C_F) +O(E^4)\right\} \per
      \label{T2}
      \ee
      The contribution
      $U_F(s_\pi, t, u)$ denotes the unitarity correction generated
      by the one-loop graphs which appear at order $E^4$ in the low-energy
expansion. It has
      the form
      \bearr
      U_F (s_\pi, t, u) & = &   \Delta_0 (s_\pi) +
      A_F(t) + B(t,u) \co
      \label{T3}
      \eearr
      with
      \bearr
      \Delta_0(s_\pi) & = & \frac{1}{2} (2 s_\pi - M^2_\pi) J^r_{\pi
      \pi} (s_\pi) + \frac{3s_\pi}{4} J^r_{KK} (s_\pi) +
      \frac{M^2_\pi}{2} J^r_{\eta \eta} (s_\pi) \co
      \nonumber \\
      A_F (t) & = & \frac{1}{16} \left[(14 M^2_K + 14 M^2_\pi - 19t)
      J^r_{K\pi} (t) + (2M^2_K + 2 M^2_\pi - 3t)
      J^r_{\eta K} (t) \right]
      \nonumber \\
      & + & \frac{1}{8} \left[ (3M^2_K - 7 M^2_\pi + 5 t) K_{K \pi}
      (t) + (M^2_K - 5 M^2_\pi + 3t) K_{\eta K} (t) \right]
      \nonumber \\
      & - & \frac{1}{4} \left[ 9 (L_{K\pi} (t) + L_{\eta K} (t)) +
      (3M^2_K - 3 M^2_\pi - 9 t) (M^r_{K\pi} (t) + M^r_{\eta K}
      (t))\right] \co
      \nonumber \\
      B(t,u) & = &  - \frac{1}{2} (M^2_K + M^2_\pi -t)
      J^r_{K\pi} (t) - ( t \leftrightarrow u).
      \label{T4}
      \eearr
      The loop integrals $J^r_{\pi \pi} (s_\pi), \ldots$ which occur
      in these expressions are listed in appendix \ref{loop}. The
functions
      $J^r_{PQ}$ and $M^r_{PQ}$ depend on the scale $\mu$ at which the
      loops are renormalized. The scale drops out in the expression
      for the full amplitude (see below).

      The imaginary part of  $F_\pi^{-2} \Delta_0 (s_\pi)$ contains
      the $I = 0$, $S$-wave $\pi \pi$ phase shift
      \be
      \delta^0_0 (s_\pi) = (32 \pi F^2_\pi)^{-1} (2s_\pi - M^2_\pi) \sigma_\pi
     + O(E^4) \co
      \label{T5}
      \ee
      as well as contributions from $K \bar{K}$ and $\eta \eta$
      intermediate states. The functions $A_F(t)$ and $B(t,u)$ are
      real in the physical region.

      The contribution $P_F(s_\pi, t, u)$ is a polynomial in
      $s_\pi, t, u$ obtained from the tree graphs at order $E^4$. We
      find
      \be
      P_F (s_\pi, t, u) =  \sum^{9}_{i=1}
      p_{i,F} (s_\pi, t, u) L^r_i \co
      \label{T6}
      \ee
      where
      \bearr
      p_{1,F} & = & 32 (s_\pi - 2M^2_\pi ) \co
      \nonumber \\
      p_{2,F} & = &8 (M_K^2 + s_\pi
      - s_l) \co
      \nonumber \\
      p_{3,F} & = & 4(M^2_K -3M_\pi^2 +2 s_\pi -t) \co
      \nonumber \\
      p_{4,F} & = & 32 M^2_\pi \co
      \nonumber \\
      p_{5,F} & = & 4 M^2_\pi \co
      \nonumber \\
      p_{9,F} & = & 2 s_l \per
      \label{T7}
      \eearr
      The remaining coefficients $p_{i,F}$ are zero. The quantities
      $L^r_i$ denote the renormalized coupling constants
which parametrize the effective lagrangian at order $E^4$ \cite{glnpb}. Their
scale dependence is
\be
L_i^r(\mu_2) =L_i^r(\mu_1) + \frac{\Gamma_i}{16\pi^2} \ln\frac{\mu_1}{\mu_2} \;
\; . \label{eq:scale}
\ee
Observable quantities are independent of the scale $\mu$, once the loop
contributions are included. The coefficients $\Gamma_i$ are displayed in table
 1, together with the value \cite{glnpb} of the couplings $L_i^r$ at
$\mu=M_\rho$.

\begin{table}[t]
\refstepcounter{table}
{Table 1: Phenomenological values and source for the renormalized coupling
constants $L^r_i(M_\rho)$ according to  Ref. \protect{\cite{glnpb}}.
The quantities $\Gamma_i$
in the fourth column determine the scale dependence of the $L^r_i(\mu)$
according to Eq. (\protect\ref{eq:scale}). $L_{11}^r$ and $L_{12}^r$ are not
directly accessible to experiment.} \label{tab:Li}
\begin{center}
\vspace{.5cm}
\begin{tabular}{|c||r|l|r|}  \hline
$i$ & $L^r_i(M_\rho) \times 10^3$ & source & $\Gamma_i$ \\ \hline
  1  & 0.7 $\pm$ 0.5 & $\pi\pi$ $D$-waves, Zweig rule & 3/32  \\
  2  & 1.3 $\pm$ 0.7 &  $\pi\pi$ $D$-waves &  3/16  \\
  3  & $-$4.4 $\pm$ 2.5 &$\pi\pi$ $D$-waves, Zweig rule&  0     \\
  4  & $-$0.3 $\pm$ 0.5 & Zweig rule &  1/8  \\
  5  & 1.4 $\pm$ 0.5  & $F_K:F_\pi$ & 3/8  \\
  6  & $-$0.2 $\pm$ 0.3 & Zweig rule &  11/144  \\
  7  & $-$0.4 $\pm$ 0.2 &Gell-Mann-Okubo,$L_5,L_8$ & 0             \\
  8  & 0.9 $\pm$ 0.3 & \small{$M_{K^0}-M_{K^+},L_5,$}&
5/48 \\
     &               &   \small{ $(2m_s-m_u-m_d):(m_d-m_u)$}       & \\
 9  & 6.9 $\pm$ 0.7 & $<r^2>_{em}^\pi$ & 1/4  \\
 10  &$ -5.5\pm 0.7$& $\pi \rightarrow e \nu\gamma$  &  $-$ 1/4  \\
\hline
11   &               &                                & $-$1/8 \\
12   &               &                                & 5/24 \\
\hline
\end{tabular}
\end{center}
\end{table}

      The contributions $C_F$  contain
      logarithmic terms, independent of $s_\pi, t$ and $u$:
      \bearr
      C_F & = & \frac{1}{256 \pi^2} \left[ 5 M^2_\pi \ln
      \frac{M^2_\pi}{\mu^2} - 2 M^2_K \ln \frac{M^2_K}{\mu^2} - 3
      M^2_\eta \ln \frac{M^2_\eta}{\mu^2} \right] \per
      \label{T8}
      \eearr
      The corresponding decomposition of the form factor $G$
      is
      \bearr
      G(s_\pi, t, u)  &=&  \frac{M_K}{\sqrt{2} F_\pi} \left\{ 1 +
      \frac{1}{F_\pi^2}(U_G +P_G +C_G) +O(E^4) \right\}\co\nn
      U_G (s_\pi, t, u) & = &  \Delta_1 (s_\pi) +
      A_G(t) + B(t,u) \co
      \label{T10}
      \eearr
      with
      \bearr
      \Delta_1 (s_\pi) & = & 2 s_\pi \left\{ M^r_{\pi \pi} (s_\pi)+
      \frac{1}{2} M^r_{KK}(s_\pi) \right\} \co
      \nonumber \\
      A_G(t) & = & \frac{1}{16} \left[ (2M^2_K + 2M^2_\pi + 3t)
      J^r_{K\pi}(t) - (2M^2_K + 2M^2_\pi -3 t) J^r_{\eta
      K}(t) \right]
      \nonumber \\
      & + & \frac{1}{8} \left[ (-3 M^2_K + 7 M^2_\pi - 5 t) K_{K \pi}
      (t) + (-M^2_K + 5 M^2_\pi - 3 t) K_{\eta K} (t) \right]
      \nonumber \\
      & - & \frac{3}{4} \left[ L_{K\pi}(t) + L_{\eta K} (t) - (M^2_K -
      M^2_\pi + t) (M^r_{K\pi} (t) + M^r_{\eta K} (t)) \right]
      \nonumber \per \\
      \label{T11}
      \eearr
      The imaginary part of $F_\pi^{-2} \Delta_1(s_\pi)$ contains the
      $I=1$, $P$-wave phase shift
      \be
      \delta^1_1 (s_\pi) = (96 \pi F_\pi^2)^{-1} s_\pi
      \sigma_\pi^3 +  O(E^4) \per
      \label{T12}
      \ee
      as well as contributions from $K\bar{K}$ intermediate states.
      The function $A_G$ is real in the physical region.

      The polynomials
      \be
      P_G =  \sum^{9}_{i=1} p_{i,G} (s_\pi,
      t, u) L^r_i
      \label{T13}
      \ee
      are
      \bearr
      p_{2,G} & = & 8(t - u) \co
\nonumber \\
      p_{3,G} & = & 4(t-M_K^2 -M_\pi^2) \co
      \nonumber \\
      p_{5,G} & = & 4M^2_\pi \co
      \nonumber \\
      p_{9,G} & = &  2 s_l \co
      \nonumber \\
      \label{T14}
      \eearr
      The remaining $p_{i,G}$ vanish. The logarithms contained in
      $C_G$ are
      \be
      C_G = - C_F.
      \label{T15}
      \ee

The form factor $R$ contains a pole part $Z(s_\pi,t,u)/(s_l-M_K^2)$ and a
regular piece $Q$. [Since the axial current acts as an interpolating
field for a kaon, the residue of the pole part is related to
the $KK\rightarrow \pi\pi$ amplitude in the standard manner.] We write
 \bearr
 R &=& \frac{M_K}{2\sqrt{2}F_\pi}\left\{\frac{Z}{s_l-M_K^2} +Q +O(E^4)\right\}
\co
\nonumber
\\ I &=& B_I +\frac{1}{F_\pi^2}(U_I +P_I +C_I) \; , \; \; I=Z,Q \; \; .
\eearr
The Born terms $B_I$ are \cite{weinkl4}
\bearr
B_Z &=& s_\pi +\nu \nonumber \co \\
B_Q&=&1.
\eearr
For the loop corrections $U_I,P_I$ and $C_I$ we find for the residue $Z$
\bearr
U_Z &=& s_\pi \Delta_0(s_\pi) + \nu \Delta_1(s_\pi) -\frac{4}{9}M_K^2 M_\pi^2
J^r_{\eta \eta}(s_\pi)
\nonumber \\
&+& \frac{1}{32}\left[ 11(s_\pi-\nu)^2 -20\Sigma (s_\pi -\nu) +12
{\Sigma}^2\right] J^r_{K\pi}(t)
\nonumber \\
&+& \frac{1}{96} \left[ 3(s_\pi -\nu) -2\Sigma\right]^2 J^r_{\eta K}(t)
\nonumber \\
&+& \frac{1}{4}(s_\pi +\nu)^2 J^r_{K \pi}(u)
\nonumber \\
&+& \frac{1}{4}(M_K^2 -M_\pi^2)\left[5(s_\pi -\nu) -6\Sigma\right] K_{K\pi}(t)
\nonumber \\
&+& \frac{1}{4}(M_K^2 -M_\pi^2)\left[3(s_\pi -\nu) -2\Sigma\right] K_{\eta
 K}(t)
\nonumber \\
&+& \frac{3}{8} \left[2s_\pi (\nu +4\Sigma) -3 s_\pi^2 + \nu^2 -16 M_\pi^2
M_K^2\right] \left[M^r_{K \pi}(t) + M^r_{\eta K} (t)\right]
\nonumber \\
&-& \frac{3}{4} (3 s_\pi +\nu -2 \Sigma)(L_{\eta K}(t) +L_{K \pi}(t)) \co
\eearr
and
\bearr
      P_Z (s_\pi, t, u) &=&  \sum^{9}_{i=1}
      p_{i,Z} (s_\pi, t, u) L^r_i \co
\eearr
with
\bearr
p_{1,Z} &=& 32 (s_\pi -2M_K^2)(s_\pi -2M_\pi^2) \co
\nonumber \\
p_{2,Z}&=& 8 (s_\pi^2 +\nu^2) \co
\nonumber \\
p_{3,Z}&=&-2\left[2(\nu +4\Sigma)s_\pi -5 s_\pi^2 -\nu^2
-16M_K^2M_\pi^2\right] \co \nonumber \\
p_{4,Z}&=& 32 \left[ \Sigma s_\pi -4M_K^2M_\pi^2\right] \co
\nonumber \\
p_{5,Z}&=&4\left[(s_\pi+\nu)\Sigma -8M_K^2M_\pi^2\right] \co
\nonumber \\
p_{6,Z} &=& 128 M_K^2M_\pi^2 \co
\nonumber \\
p_{8,Z} &=&  64 M_K^2M_\pi^2 \per
\eearr
The remaining $p_{i,Z}$ vanish. Finally, the logarithms in $C_Z$ are
\bearr
C_Z&=& -\frac{M_K^2 -M_\pi^2}{128 \pi^2} \left[ 3 M^2_\pi \ln
      \frac{M^2_\pi}{\mu^2} - 2 M^2_K \ln \frac{M^2_K}{\mu^2} -
      M^2_\eta \ln \frac{M^2_\eta}{\mu^2} \right]
 \; \; .
\eearr
The nonpole part Q receives the following one-loop contributions:
\bearr
U_Q&=& \Delta_0(s_\pi) +\frac{M_K^2-s_l}{32}\left\{11 J^r_{K \pi}(t) +8J^r_{K
\pi}(u) +3 J^r_{\eta K}(t)\right\}
\nonumber \\
&-& \frac{1}{8}(5(s_\pi -\nu) +5(M_K^2-s_l)-6\Sigma)K_{K\pi}(t)
\nonumber \\
&-& \frac{1}{8}(3(s_\pi -\nu) +3(M_K^2-s_l)-2\Sigma)K_{\eta K}(t)
\nonumber \\
&-&\frac{9}{4}(L_{\eta K}(t) + L_{K \pi}(t))
\nonumber \\
&+&\frac{3}{8}(4(\nu +2M_\pi^2) -3(M_K^2 -s_l))(M^r_{K \pi}(t) + M^r_{\eta
K}(t)),
\eearr
and
\bearr
      P_Q (s_\pi, t, u) &=&  \sum^{9}_{i=1}
      p_{i,Q} (s_\pi, t, u) L^r_i \co
\eearr
with
\bearr
p_{1,Q}&=& 32(s_\pi -2M_\pi^2) \co
\nonumber \\
p_{2,Q}&=& 8(M_K^2 -s_l) \co
\nonumber \\
p_{3,Q}&=&2( 4(s_\pi -2M_\pi^2) +M_K^2-s_l) \co
\nonumber \\
p_{4,Q}&=& 32M_\pi^2 \co
\nonumber \\
p_{5,Q}&=& 4 \Sigma \co
\nonumber \\
p_{9,Q}&=& 2\left[(s_\pi +\nu) -(M_K^2 -s_l)\right]\per
\eearr
The remaining $p_{i,Q}$ vanish. Finally, the logarithms in $C_Q$ are
\bearr
C_Q&=&   \frac{1}{128 \pi^2} \left[ 5 M^2_\pi \ln
      \frac{M^2_\pi}{\mu^2} - 2 M^2_K \ln \frac{M^2_K}{\mu^2} - 3
      M^2_\eta \ln \frac{M^2_\eta}{\mu^2} \right]
 \; \; .
\eearr

The first nonvanishing contribution in the chiral expansion of the form factor
$H$ is   due
to the chiral anomaly \cite{wessz}.
 The prediction is  \cite{wesszkl4}
 \be
      H = -\frac{\sqrt{2} M^3_K}{8 \pi^2 F^3_\pi} = -2.66 \co
      \label{T16}
      \ee
      in excellent agreement with the experimental value \cite{ross} $H=-2.68
\pm 0.68$. The
next-to-leading order corrections to $H$ have also been calculated
\cite{llamet}.
 If the new low-energy parameters are
estimated using the vector-mesons only, these corrections are small.

      The results for $F,G$ and $R$ must satisfy two nontrivial
      constraints: i) Unitarity requires that $F,G$ and $R$ contain, in
      the physical region $4M^2_\pi \leq s_\pi \leq (M_K-m_l)^2$, imaginary
      parts governed by $S$- and $P$-wave $\pi \pi$ scattering [these
      imaginary parts are contained in the functions
      $\Delta_0(s_\pi), \Delta_1(s_\pi)]$. ii) The scale dependence
      of the low-energy constants $L^r_i$ must be compensated  by
      the scale dependence of $U_{F,G,Z,Q}$ and $C_{F,G,Z,Q}$ for all values
      of
      $s_\pi, t, u, M^2_\pi, M^2_K$. [Since we work at order $E^4$ in the
chiral expansion, the
      meson masses appearing in the above expressions satisfy the
      Gell-Mann-Okubo mass formula.] We have checked that these
      constraints are satisfied.

It is one of the aims of this article to determine the low-energy constants
$L_1^r,L_2^r$ and $L_3$ from  experimental data on $K^+\rightarrow
\pi^+\pi^-e^+\nu_e$ decays and on $\pi\pi$ threshold parameters. In Ref.
\cite{bijnenskl4,riggen}, the above one-loop expressions  have been used for
this purpose.
Because the one-loop contributions are rather large, the values of the
$L_i$'s so extracted suffer from substantial uncertainties. In the following
section, we therefore
first estimate the effects from higher orders in
the chiral expansion, using then this improved representation for the form
factors in a comparison with the data.

\setcounter{equation}{0}
\setcounter{subsection}{0}
\section{Beyond one-loop}

To investigate the importance of higher order terms, we employ the method
developed in Ref. \cite{dgl}. It amounts to writing a dispersive
representation of the partial wave amplitudes, fixing the subtraction
constants using chiral perturbation theory. Here, we estimate the
higher order terms in the $S$-wave projection of the amplitude $F_1$,
\be
f(s_\pi,s_l) = (4\pi X)^{-1}\int d\Omega F_1(s_\pi,t,s_l)\co
\label{un2}
\ee
 because
this form factor plays a decisive role in the determination of $L_1^r,L_2^r$
and $L_3$, and it is influenced by $S$-wave $\pi\pi$ scattering which is known
 \cite{truongswave,truongkl4} to produce substantial corrections.

\subsection{Analytic properties of $f(s_\pi,s_l$)}

Only the crossing even part
\be
F_1^+=XF^+ +\sigma_\pi (PL)\cos \thp \cdot G^-
\ee
contributes in the projection (\ref{un2}).
 The partial wave $f$ has the following analytic properties:
\begin{enumerate}
\item
At fixed $s_l$, it is analytic in the complex $\spi$-plane, cut along the real
axis for Re $\spi \geq 4M_\pi^2$ and  Re $\spi \leq 0$.
\item
In the interval $0\leq s_\pi \leq 4M_\pi^2$, it is real.
\item
In $ 4M_\pi^2\leq s_\pi \leq 16 M_\pi^2$, its phase  coincides with the
isospin zero $S$-wave phase $\delta_0^0$ in elastic $\pi\pi$ scattering,
\be
f_+ = e^{2i\delta_0^0}f_- \co f_\pm = f(s_\pi \pm i\epsilon,s_l).
\label{un3a}
\ee
\end{enumerate}
The proof of these properties is standard \cite{spearman}. Here we only note
that the presence of the cut for $s_\pi \leq 0$ follows from the relations
\bearr
t=M_\pi^2 +\frac{M_K^2 +s_l - s_\pi}{2} -\sigma_\pi X \cos \theta _\pi \co \nn
\label{un4}
t(\cos \theta_\pi=-1,s_\pi < 0) \geq (M_K+M_\pi)^2.
\label{bo1}
\eearr
Since $F^+$ and $G^-$ have cuts at $t\geq (M_K+M_\pi)^2$ [see e.g. Eqs.
\ref{T2}--\ref{T11}], the claim is proven.

\subsection{Unitarization}

We introduce the Omn\`{e}s-function
\be
\Omega(s_\pi) = \exp \left[ \frac{\spi}{\pi}\int_{4M_\pi^2}^{\Lambda^2}
\frac{ds}{s}\frac{\delta_0^0(s)}{s-\spi}\right] \co
\ee
where $\Lambda$ will be chosen of the order of 1 GeV below.
According to (\ref{un3a}), multiplication  by $\Omega^{-1}$
removes the cut in $f$ for  $4M_\pi^2 \leq \spi\leq 16 M_\pi^2$.
Consider now
\be
f=f_L+f_R \co
\ee
where $f_L(f_R)$ has only the left-hand (right-hand) cut, and introduce
\be
v=\Omega^{-1} (f -f_L) \; .
\ee
Then $v$ has only a right-hand cut, and we may represent it in a dispersive
way,
\be
v=v_0 +v_1 \spi+\frac{\spi^2}{\pi} \int_{4M_\pi^2}^\infty
\frac{ds}{s^2}\frac{{\mbox{Im}}\Omega^{-1}(f-f_L)}{s-\spi} \; .
\ee
We expect the contributions from the integral beyond 1GeV$^2$ to be small.
Furthermore, $\Omega^{-1}f$ is approximately real between $16M_\pi^2$ and
1GeV$^2$, as a result of which one has
\be
v = v_0 + v_1 \spi -\frac{\spi^2}{\pi}\int_{4M_\pi^2}^{\Lambda^2}
 \frac{ds}{s^2}\frac{f_L{\mbox{Im}}\Omega^{-1}}{s-\spi} \per
\ee
For given $v_0,v_1,f_L$ and $\Omega$, the form factor $f$ is finally obtained
from
\be
f=f_L + \Omega v \per
\label{un5}
\ee
The behaviour of $f_L$ at  $\spi \rightarrow 0$ is
governed by the large $|t|$-behaviour of $F^+$ and $G^-$, see (\ref{bo1}).
Therefore, instead
of using CHPT to model $f_L$, we approximate the left-hand cut by resonance
exchange. To pin down the
subtraction constants $v_0$ and
$v_1$, we require that the threshold expansion of $f$ and $f_\CHPT$
agree
up to and including terms of order $O(E^2)$. For a specific choice of $f_L$,
this fixes ${v_0,v_1}$ in terms of the quantities which occur in the one-loop
representation of $F^+$ and $G^-$.
In the case where $f_L=0$, $f$ has then a particularly simple form at $s_l=0$,
\bea
f(s_\pi,s_l=0)|_{f_L=0} &=& \Omega(v_0 + v_1 s_{\pi}) \co \nn
v_0 &=& \frac{M_K}{\sqrt{2} F_\pi}\biggr\{ 1.05 +\frac{1}{F_\pi^2}
\biggr[-64 M_\pi^2 L^r_1 + 8 M_K^2 L^r_2 \nn
  & &  +2(M_K^2-8M_\pi^2)L_3
  +\frac{2}{3}(M_K^2-4M_\pi^2)(4L^r_2+L_3)\biggr]\biggr\} \co \nn
v_1 &=& \frac{M_K}{\sqrt{2} F_\pi}\biggr\{0.38+\frac{1}{F_\pi^2}
 \biggr[32L^r_1+8L^r_2+10L_3 \nn
 & & \qquad \qquad -\frac{2}{3}\frac{M_K^2-4M_\pi^2}{4M_\pi^2}(4L^r_2+L_3)
\biggr]\biggr\}\per
\label{un6}
\eea
We relegate the details of the calculation of $f_L,v_0$ and of $v_1$ to
 appendix
 \ref{calcv0}.

In the partial wave $f$, the effects of the final-state interactions are
substantial, because they are related to the $I=0,S$-wave $\pi \pi$ phase
shift. On the other hand, for the leading partial wave in
$G^+=g e^{{i\delta_p}}+\cdots$, these effects are reduced, because the phase
$\delta_p$ is small at low energies. We find it more difficult to assess an
estimate for the higher order corrections in this case -- we come back to this
point in the following section.

\setcounter{section}{4}
\setcounter{equation}{0}
\setcounter{subsection}{0}
\section{Determination of $L_1,L_2$ and $L_3$}

Here we determine the low-energy constants $L^r_1,L^r_2$
and $L_3$ from experimental data on $K^+\rightarrow \pi^+\pi^-e^+\nu_e$
decays and on $\pi\pi \rightarrow \pi\pi$ threshold parameters, using the
improved $S$-wave amplitude $f$ set up above. We are  aware that our
results will not be the last word: future kaon facilities like DA$\Phi$NE
\cite{dafne} will
allow a more refined comparison of the chiral representation with the data.
Nevertheless, we believe that it is instructive to see what one has to expect
from higher order contributions. A
comparison with earlier work \cite{riggen} will be provided at
the end of this section.

\subsection{The data}

Experimentally
the study of $K_{l4}$ decays is dominated by the work of Rosselet
et al. \cite{ross} which measures $K^+\rightarrow \pi^+\pi^-e^+\nu_e$ with
good statistics.
The total decay rate, the
absolute value of the form factors $F,G$ and of $H$  and the difference of the
phases
$\delta_0^0 - \delta_1^1$ were determined by use of
\bea
F &=& f_s e^{i\delta_s} + f_p e^{i\delta_p}\cos \thp + D-\wave \co \nn
G &=& g e^{i\delta_p} + D-\wave \co \nn
H &=& h e^{i\delta_h} + D-\wave \per
\eea
The form factor $f_p$ was found to be compatible with zero and hence set equal
to zero when the final value for $g$ was derived. No $s_\pi,s_l$ dependence of
the ratios $g/f_s$ and $h/f_s$ was seen. Parametrizing $f_s$ in the form
\bearr
f_s &=& f_s(0) (1 + \lambda_f q^2) \co \nn
q^2&=&(s_\pi-4M_\pi^2)/4M_\pi^2\co
\label{fl1}
\eearr
then gives
\bearr
g &=& g(0) (1 + \lambda_g q^2) \co \nn
h &=& h(0) (1 + \lambda_h q^2) \co \; \;
\eearr
with
$\lambda_f = \lambda_g = \lambda_h$. Rosselet et al. found \cite{ross}
\bea
\begin{array}{rcr}
f_s(0)& =& 5.59 \pm 0.14 \co \\
g(0)& =& 4.77 \pm 0.27 \co \\
h(0)& =& -2.68 \pm 0.68 \co \\
\lambda_f&=&0.08\pm0.02 \co
\end{array}
\label{dl1}
\eea
where we have used $|V_{us}| =0.22$ in transcribing their results.
Notice that the experimental numbers (\ref{dl1})
have been obtained in \cite{ross} under assumptions which are in conflict with
our theoretical
formulae, like absence of higher waves, $s_l$ independence of the form factors
and equality of the slopes, $\lambda_f=\lambda_g=\lambda_h$. It would of
course be
desirable to analyze forthcoming data without any additional assumptions.

 The total decay rate  is  \cite{ross}
\bearr \Gamma_{K_{e4}} =
(3.26 \pm 0.15)
10^3 \mbox{sec}^{-1} \per
\label{dl2}
\eearr
In
\cite{riggen} it has
been observed that, using as input the central values (\ref{dl1}),
 one obtains
$\Gamma_{K_{e4}}=2.94\cdot 10^3$ sec$^{-1}$, which disagrees with (\ref{dl2}).
On the other hand,
the  value (\ref{dl2}) was used in Ref. \cite{ross} to normalize the form
factors in (\ref{dl1}). We do not understand the origin of this contradiction.

The $\pi\pi$ threshold data used below are taken from Ref.
\cite{nagels}. We display them in table 2, column 6.

\subsection{The fits}

In the following, we perform various fits
 to $f_s(0),\lambda_f,g(0)$ and to the $\pi\pi$ threshold
parameters listed in table 2. We introduce for this purpose the
quantities
 \bea\label{proj} \fff(s_{\pi},s_l) &=&\left| (4\pi X)^{-1} \int
d\Omega F_1(s_{\pi},t,
s_l) \right|=\left|f(s_\pi,s_l)\right|  \co \nn
\ffg(s_{\pi},s_l) &=& \left| \frac{3}{8 \pi} \int d\Omega \sin^2\thp
G(s_{\pi},t,s_l) \right|\co
\eea
where the factor $3/2\sin^2\theta$ appears because $G$ is expanded
in derivatives of  Legendre polynomials.
Below, we confront [$f_s(0)$, $g(0)$] with [$\bar{f}(4M_\pi^2,s_l)$],
 $\bar{g}(4M_\pi^2,s_l)$], which depend on $s_l$. Furthermore, we compare
the slope $\lambda_f$ with
\be
\bar{\lambda}_f(\spi,s_l)=
\frac{\fff(\spi,s_l)-\fff(4M_\pi^2,sl)}{\fff(4M_\pi^2,s_l)}\frac{4M_\pi^2}
{s_\pi-4M_\pi^2} \co
 \ee
which depends on both $s_\pi$ and $s_l$. Below we use these dependences to
estimate systematic uncertainties in the determination of the low-energy
couplings. [In future high statistics experiments, the $s_l$-dependence of the
form factors will presumably  be resolved. It will  be easy to adapt the
procedure to this case.]

We have used MINUIT \cite{minuit} to perform the fits. The results for the
choice $s_l=0,s_\pi=4.4M_\pi^2$ are given in table 2.
\begin{table}[t]
{Table 2:
Results of  fits with  one-loop and unitarized  form factors, respectively. The
errors quoted for the $L^r_i$'s are statistical only. The $L^r_i$ are given in
units of $10^{-3}$ at the scale $\mu=M_\rho$, the scattering lengths $a^I_l$
and the slopes $b_l^I$ in appropriate powers of $M_{\pi^+}$.}
\begin{center}
\vspace{.5cm}
\begin{tabular}{|c|cc|cc|c|}
\hline
         & \multicolumn{2}{c|}{ $K_{e4}$ data
alone}&\multicolumn{2}{c|}{$K_{e4}$  and $ \pi \pi$ data} & experiment\\
         & one-loop        & unitarized & one-loop & unitarized
&\cite{nagels} \\ \hline
$L^r_1$& $0.65 \pm 0.27$&$0.36\pm0.26$  & $0.60\pm0.24$  &$0.37\pm0.23$ & \\
$L^r_2$& $1.63 \pm 0.28$ &$1.35\pm0.27$ &$1.50\pm0.23$    &$1.35\pm0.23$
&\\ $L_3$&$-3.4\pm1.0$   &$-3.4\pm1.0$   &$-3.3\pm0.86$ &$-3.5\pm0.85$  & \\
\hline

$ a_0^0$& $0.20$& $0.20$ &$ 0.20$ & $0.20$       &$0.26\pm0.05$\\

 $b_0^0$ & $0.26$ & $0.25$& $0.26$ & $0.25$&$0.25\pm0.03$\\

$-$10 $a_0^2$ &$0.40$ &$0.41$ &$0.40$ &$0.41$ &$0.28\pm0.12$\\

$-$10 $b_0^2$     &$0.67$ &$0.72$ &$0.68$ &$0.72$ &$0.82\pm0.08$\\

$10 a_1^1$     &$0.36$  &$0.37$  &$0.36$  &$0.37$  &$0.38\pm0.02$\\

$10^2 b_1^1$     &$0.44 $ &$0.47$  &$0.43$  &$0.48$  &\\

$10^2 a_2^0$     &$0.22 $  &$0.18$   &$0.21$   &$0.18$   &$0.17\pm0.03$ \\

$10^3 a_2^2$     &$0.39 $   &$0.21 $   &$0.37 $   &$0.20 $   &$0.13\pm0.3$\\
\hline
$\chi^2/N_\mtiny{DOF}$& $0/0$&$0/0$&$8.8/7$&$4.9/7$&  \\
\hline
\end{tabular}
\end{center}
\end{table}
In the columns denoted by "one-loop", we have evaluated $\bar{f},\bar{g}$ and
 $\bar{\lambda}_f$ from the one-loop representation given above\footnote{We
always use for  $L_4^r,\ldots,L_9^r$ the values quoted
in table 1.}. In the fit with the unitarized form
factor  (columns 3 and 5), we have evaluated $\bar{f}$ from
 Eqs. (\ref{un5},\ref{fvs}), inserting in the Omn\`{e}s function the
parametrization of the $\pi \pi$ $S$-wave phase shift proposed by Schenk
\cite[solution B]{schenk}. For the form factor $G$,
we have again used the one-loop representation.
 The statistical errors
quoted for the $L_i$'s are the ones generated by the procedure MINOS in MINUIT
and correspond to an increase of $\chi^2$ by one unit.

A few  remarks are in order at this place.
\begin{enumerate}
\item
 It is seen that the
overall description of the $\pi \pi$ scattering data is better using
the unitarized form factors, in particular so for
the $D$-wave scattering lengths.
\item
The errors quoted do not give account of the fact that the simultaneous
determination of the three constants produces a strong correlation between
them.
To illustrate this point we note that, while
the values of the $L_i$'s in
column 4 and 5 are apparently consistent with each other within one error bar,
the $\chi^2$ in column 5 increases from 4.9 to 30.7 if
the $L_i$'s from column 4 are used in the evaluation of $\chi^2$ in
column 5. (For a discussion about the interpretation of the errors see
\cite{minuit}).
\item
The low-energy constants $\bar{l}_1,\bar{l}_2$ which occur in $SU(2)_L\times
SU(2)_R$ analyses may be evaluated from a given set of $L_1^r,L_2^r$ and $L_3$
 \cite{glnpb}. Their value changes in a significant way by
using the unitarized amplitude instead of the one-loop formulae: the
 values for $(\bar{l}_1,\bar{l}_2)$ in column 4 and 5 are
$(-0.5\pm 0.88,6.4\pm 0.44)$ and $(-1.7\pm 0.85, 6.0\pm 0.4)$, respectively.

\item
 $L_1^r,L_2^r$ and $L_3$ are related to  $\pi \pi$
phase-shifts through sum rules \cite{truongsum,sternsum}. In principle, one
should
take these constraints into account as well\footnote{We thank B. Moussallam for
pointing this out to us.}. We do not consider them here, because we find it
very difficult to assess a reliable error for the integrals over the
total $\pi\pi$ cross sections which occur in those relations.
\end{enumerate}

The {\it{statistical}} error in the data is only one source of the uncertainty
in the
low-energy constants, which are in addition affected by the ambiguities in the
estimate of the higher order corrections. These {\it systematic}
uncertainties have several sources:
\begin{enumerate}
\item[i)]
Higher order corrections to $\ffg$ have not been taken into account.
\item[ii)]
The determination of the contribution from the left-hand cut is not unique.

\item[iii)]
The quantities $\fff$ and $\ffg$ depend on $s_l$, and $\bar{\lambda}_f$ is a
function of both $s_l$ and $s_\pi$.
\item[iv)]
The Omn\`{e}s function depends on the
elastic $\pi\pi$ phase shift and on the cutoff $\Lambda$ used.
\end{enumerate}

We have
considered carefully these effects.
As for the first point, we have evaluated the higher orders in $\bar{g}$ in
two ways: \begin{itemize}
\item
We define the quantity \cite{riggen}
 \be
\Delta \ffg = \frac{(g(0)-\ffg^{(2)})^2}{\ffg^{(2)} } \co
\ee
where $\ffg^{(2)}$ is the CHPT prediction at leading-order. We then add $\Delta
\ffg$ in quadrature to the experimental error in $g(0)$ and redo the fit.
This generates slightly larger errors than before. To illustrate, the entries
$(0.23,0.23,0.85)$ in column 5 in table 2 become $(0.29,0.28,1.1)$.

\item
The main contribution to the one-loop correction in $\ffg$ stems from $L_3$.
On the other hand,
the low-energy constants $L^r_i$ are saturated by resonance exchanges whose
contribution
is evaluated in the limit of large resonance mass \cite{eck}.
We find it therefore resonable to estimate higher orders by using the complete
resonance propagators.  In order to evaluate the uncertainty induced by this
 correction, we made the fit including the full propagators in
$\ffg$ and in $\fff$.
This changes the central values for the $L_i$'s, see table 3.
\end{itemize}
\begin{table}[t]
{Table 3:
Fits made with different theoretical input to determine the systematic
uncertainties in  $L_1^r,L_2^r$ and $L_3$.
 $\Delta L_1$ shows the difference to the value of  $L^r_1$
displayed in the second row.
  $L^r_i$
are given in units of $10^{-3}$ at the scale $\mu=M_\rho$. The fits include
$K_{e4}$ and $\pi\pi$ data. }
\begin{center}
\vskip 0.5cm
\begin{tabular}{|c|lr|lr|lr|}
\hline
 & $ L^r_1$&$ \Delta L_1$ &$  L^r_2$&$ \Delta L_2$ & $
L_3$&$ \Delta L_3$\\ \hline
{ $5^\mtiny{th}$ column table 2}& $0.37$ & & $1.35$& & $-3.5$& \\
\hline

{\small full propagators} & $0.34$ & $-0.03$&$1.08$
&$-0.27$&$-2.8$&$-0.7$\\
$f_L = 0$ & $0.42$&$0.05$ & $1.30$ &$-0.05$&$-3.5$&$0.0$\\

$s_\pi = 5.6 M_\pi^2,s_l=0$ & $0.38$ & $0.01$& $1.35$ & $0.00$ & $-3.5$ &$
0.0$ \\

$s_\pi=4.4M_\pi^2,s_l = M_K^2/10$ &$ 0.33$ &$ -0.04$ & $1.26$ & $-0.09$ &
$-3.3$ & $-0.2$ \\
$\pi \pi$ phase from \cite[solution A]{schenk} &$ 0.33$ &$ -0.04$ & $1.35$ &
$0.00$ & $-3.5$ & $0.0$ \\
$\pi \pi$ phase from \cite[solution C]{schenk} &$ 0.46$ &$ 0.09$ & $1.36$ &
$0.01$ & $-3.5$ & $0.0$ \\ \hline
\end{tabular}
\end{center}
\end{table}
Concerning the effect of the left-hand cut, we estimate its uncertainty by
dropping this piece altogether.
The results of the fit in this case  are  again given  in table 3.
It is seen that the
$L^r_i$'s depend rather weakly on the presence of the left-hand cut. Figures 2
and 3 illustrate the effect of $f_L$ in more detail. In Fig. 2 is shown
the
form factor $\fff$  with and without the left-hand cut
(dashed and dash-dotted line, respectively). In both cases, $L_1^r,L_2^r,L_3$
from (\ref{dl3}) have been used.
 At large values of $s_\pi$, the difference between the two evaluations of
 $\bar{f}$ is
not negligible  for
the following reason. At threshold, the value of the form factor and of its
slope cannot change much by construction--these quantities are
constrained in our procedure by the low-energy expansion at one-loop order. On
the other hand,
the quadratic piece is unconstrained and receives contributions
from $f_L$ through the dispersive integral. The corresponding change in $\fff$
is less than $10\%$ at $s_\pi=10M_\pi^2$.
 In Fig. 3, we display
the differential decay rate $d\Gamma/ds_\pi$, obtained by using the form factor
with
and without $f_L$. The figure illustrates that the contribution from the
quadratic piece in $s_\pi$ to observables is strongly suppressed by phase
space and will therefore be difficult to observe.

The dependence of the fits on $s_l,s_\pi$ and on  the $\pi \pi$ phase
shift used
in the Omn\`{e}s function $\Omega$  is illustrated in the  last
four rows in table 3. Furthermore, changing the cutoff $\Lambda=1$GeV in
$\Omega$ to $\Lambda=0.8$GeV induces small changes in the $L_i$'s only. We
conclude that a global fit  to all the available
data  is  rather stable against the systematic uncertainties considered here.

To finally give the best determinations of $L^r_1,L^r_2$ and $L_3$, we
take the central values from the global fit displayed in table 2, column 5. For
the corresponding errors, we take the ones generated by using the theoretical
error bars for the higher orders in $\bar{g}$, and find
 in this manner
\bearr
\begin{array}{rrr}
10^3L^r_1(M_\rho)&=& 0.4 \pm 0.3 \co\\
10^3L^r_2(M_\rho)&=&1.35\pm0.3 \co\\
10^3L_3(M_\rho)&=&-3.5 \pm 1.1\per
\label{dl3}
\end{array}
\eearr
For $ SU(2)_L\times SU(2)_R$ analyses it is useful to know the corresponding
values for the constants $\bar{l}_1$ and $\bar{l}_2$,
\bearr
\begin{array}{rrr}
\bar{l}_1&=&-1.7 \pm 1.0  \co\\
\bar{l}_2&=&6.1 \pm 0.5 \per
\label{dl4}
\end{array}
\eearr

The value and uncertainties in these couplings play a decisive role in a
planned experiment \cite{nemenov} to measure the lifetime of $\pi^+ \pi^-$
atoms,
which will provide a completely independent measurement of the $\pi \pi$
scattering lengths $|a_0^0-a_0^2|$.

One motivation for the analysis in \cite{bijnenskl4,riggen} was to test the
large $N_C$
prediction $L_2^r=2L_1^r$. The above result shows that a small non-zero
value is preferred.
 To obtain a clean  error analysis, we have repeated the fitting
procedure using the variables
\bea
X_1&=&L_2^r-2L_1^r-L_3\co \nn
X_2^r&=&L_2^r\co\nn
X_3&=&(L_2^r-2L_1^r)/L_3 \nonumber \per
\eea
We
performed a fit to $K_{e4}$ and $\pi\pi$ data, including the theoretical error
in $G$ as discussed above, and found
\bearr
X_1&=&(4.8 \pm 0.8)\cdot 10^{-3} \co \nn
X_3&=&-0.17^{+0.12}_{-0.22}\per
\label{dl6}
\eearr
The result is that the large $N_C$ prediction works remarkably well.

\subsection{Comparison with earlier work}

It is of interest to compare the present procedure to determine the low-energy
constants $L_1^r,L_2^r$ and $L_3$ with the method used in \cite{riggen}. There
are two main differences:
\begin{enumerate}
\item
The definition of the slope $\lambda_f$ and of the threshold value of the form
factors $f_s,g$ chosen in \cite{riggen} differs from the one used here.
These quantities have of course a unique meaning in principle -- on the other
hand, one may
wish to approximate a particular experimental situation. The procedure used in
\cite{riggen} was adapted to Ref.  \cite{ross}, whereas a slight variation of
the method proposed
here may be useful once the $s_l$-dependence of the form factors has
experimentally  been resolved.
\item
Higher order corrections are estimated in \cite{riggen} in a rather crude
manner. In the present approach, the final-state interactions in the
$I=0,S$-wave amplitude are instead taken into account,
and  higher order terms in $\bar{g}$ are estimated with resonance
exchange.
\end{enumerate}

The main effect of these differences can be described as follows. The different
slope and form factors used in \cite{riggen} lead to slightly different central
values for
$L_1^r,L_2^r$ and $L_3$ at one-loop order, whereas the errors turn out to be
very similar in both cases. The higher order estimates in \cite{riggen}
lead to the
same central values with  larger error bars, whereas the unitarization
performed
in the present work leads to different central values with slightly smaller
error bars than before, see columns 2/3 and 4/5 in table 2. This effect can be
easily understood by considering the simplified expression (\ref{un6}), which
shows how the Omn\`{e}s function   affects the influence
of the $L_i$'s and hence their value in the fit.

\subsection{Improvements}

As we mentioned at the beginning of this section, there is room for improvement
in the above treatment, both on the theoretical and on the experimental
side. Concerning the latter, one should determine in future experiments the
form factors $f_s$ and $g$ without additional assumptions \cite{ross} which are
in contradiction with the chiral representation. It remains to be seen
whether this can be  achieved
by  comparing the data directly with a modified chiral representation. In
the
latter, the full $S$- and $P$- wave parts of $F_1$ and $F_2$ could be
inserted,
using the chiral representation solely to describe the small background effects
due to higher partial waves $l\ge 2$.
To  be more precise, one would take for $R$ and $H$ the one-loop chiral
representation, whereas for $G$ one writes
\bearr
G&=&g(s_\pi,s_l) e^{{i\delta_p}} + \Delta G\co \nn
\Delta G&=& G_\mtiny{CHPT} -\frac{3}{8\pi}\int d\Omega \sin^2\thp
G_\mtiny{CHPT}\co
\eearr
and similarly for $F$. The unknown amplitudes $g(s_\pi,s_l),f_s(s_\pi,s_l)$
and the phases $\delta_p,\delta_s$ would then be determined from the data.
We have checked that, if the errors in the form factors determined in this
manner can be reduced by e.g. a factor 3
 with respect to the
ones shown in (\ref{dl1}), one could pin down particular combinations of
$L_1^r,L_2^r$ and $L_3$ to considerably better precision than was shown above.
This is true independently of an eventual improvement in the theoretical
determination of the higher order corrections in the form factor $G$ -- which
is a theoretical challenge in any case.

\setcounter{equation}{0} \setcounter{subsection}{0}
\section{Predictions}

Having determined the constants $L_1^r,L_2^r$ and $L_3$, there are several
predictions which we can make.
 Whereas the slope $\lambda_g$  was assumed to coincide
with the slope $\lambda_f$ in the final analysis of the data in Ref.
\cite{ross}, these two quantities may differ in the chiral representation.
Furthermore, our amplitudes allow us to evaluate partial and total decay
rates. In this section, we consider the slope $\lambda_g$ and the total rates.

\subsection{The slope $\lambda_g$}

We consider the form factor $\bar{g}$ introduced in  (\ref{proj}) and
determine its slope
 $\lambda_g$
\be
\bar{g}(s_\pi,s_l)=\bar{g}(4M_\pi^2,sl)(1+\lambda_g(s_l) q^2 +O(q^4))
\ee
from the one-loop expression for $G$. The result is
$\lambda_g(0)=0.08$. As the slope is a one-loop effect,
higher order corrections may affect its value substantially. For this reason,
we have evaluated $\lambda_g$
 also from the modified form factor obtained by using the complete
resonance propagators (and the corresponding $L_i$'s), compare the discussion
above.
The change is $\Delta \lambda_g =0.025$. We believe this to be a generous error
estimate and obtain in this manner
\be
\lambda_g(0)=0.08\pm 0.025 \per
\label{eslope}
\ee
The central value indeed agrees  with the slope $\lambda_f$ in (\ref{dl1}).

\subsection{Total rates}
Once the leading partial waves $\bar{f}$ and $\bar{g}$ are known from e.g.
$K^+\rightarrow \pi^+\pi^-e^+\nu_e$ decays, the chiral representation allows
one to predict the remaining rates within rather small uncertainties. We
illustrate the procedure for $K^+\rightarrow \pi^0\pi^0e^+\nu_e$.
According to Eq. (\ref{i2}),
 the relevant amplitude is determined by
$F^+,G^-,R^+$ and $H^-$. The contribution from $H$  is kinematically
strongly suppressed and completely negligible in all total
rates, whereas the contribution from $R$ is
 negligible in the electron modes. Using the chiral representation
of the  amplitudes $F^+$ and $G^-$, we find that the rate is
 reproduced to about 1\%,
if one neglects $G^-$ altogether and uses only the leading partial wave
in
the remaining amplitude, $F_1^+\simeq -X\bar{f}$. From the measured \cite{ross}
form factor $\bar{f}=5.59(1+0.08q^2)$ we then find $\Gamma_{K^+\rightarrow
\pi^0\pi^0e^+\nu_e}=1625$sec$^{-1}$. Finally, we estimate the error from
\bea
\Delta\Gamma&=& \left\{
[\Gamma(f_s(0)+\Delta f_s,\lambda_f)-
     \Gamma(f_s(0),\lambda_f)]^2 +\right.\nn
&&\left.[\Gamma(f_s(0),\lambda_f+\Delta \lambda_f)-
 \Gamma(f_s(0),\lambda_f)]^2
\right\}^{1/2}=90 {\mbox{sec}}^{-1}\co\nn
\eea
where $\Delta f_s=0.14,\Delta \lambda_f=0.02$.
The final result for the rate is shown in the row "final prediction" in
table 5, where we have also listed the tree and the one-loop result, together
with the experimental data. The evaluation of the remaining rates is done in a
similar manner -- see table 4 for the simplifications used and table 5 for the
corresponding predictions.
\begin{table}[t]
{Table 4: Approximations used to evaluate the total rates in table 5. Use of
$\bar{f}=\bar{f}_\mtiny{CHPT}, \bar{g}=\bar{g}_\mtiny{CHPT}$ reproduces the
one-loop results in table 5 to about $1\%$.}

\begin{center}
\vspace{.5cm}
a) $K^+$ decays

\vspace{.2cm}
\begin{tabular}{|c|c|c|c|}
\hline
& $\pi^0\pi^0 e^+ \nu_e$&$\pi^+\pi^-\mu^+
\nu_\mu$&$\pi^0\pi^0 \mu^+ \nu_\mu$
\\ \hline
&&&\\
$F_1$&$-X\bar{f}$&
$X\bar{f}+\sigma_\pi (PL)\cos\thp \bar{g}$&$-X\bar{f}$
\\
$F_2$&$0$&
$\sigma_\pi (s_\pi s_l)^{1/2}\bar{g}$&$0$
\\
$F_3$&0&0&0
\\
$F_4$&$(PL)\bar{f}$&
$-\left\{(PL)\bar{f}+s_lR_\mtiny{CHPT}+\sigma_\pi X\cos\thp \bar{g}\right\}$&
$\left\{(PL)\bar{f}+s_lR_\mtiny{CHPT}\right\}$
\\ \hline
\end{tabular}

\vspace{1cm}
b) $K^0$ decays. Shown are the amplitudes divided by $\sqrt{2}$.

\vspace{.2cm}
\begin{tabular}{|c|c|c|}
\hline
& $\pi^0\pi^-e^+ \nu_e$&$\pi^0\pi^- \mu^+ \nu_\mu$
\\ \hline
&&\\
$F_1$&$XF^-_\mtiny{CHPT}+\sigma_\pi (PL)\cos\thp \bar{g}$&
      $XF^-_\mtiny{CHPT}+\sigma_\pi (PL)\cos\thp \bar{g}$
\\
$F_2$&$\sigma_\pi (s_\pi s_l)^{1/2}\bar{g}$&
      $\sigma_\pi (s_\pi s_l)^{1/2}\bar{g}$
\\
$F_3$&0&0
\\
$F_4$&$-\left\{(PL)F^-_\mtiny{CHPT}+\sigma_\pi X\cos\thp \bar{g}\right\}$&
$      -\left\{(PL)F^-_\mtiny{CHPT} +s_lR_\mtiny{CHPT}+\sigma_\pi X\cos\thp
\bar{g}\right\}$ \\ \hline
\end{tabular}
\end{center}
\end{table}

\begin{table}[t]
{Table 5:
Total  decay rates in sec$^{-1}$. To evaluate the rates at one-loop accuracy,
we have used $L_1^r,L_2^r$ and $L_3$ from (\ref{dl3}). The remaining low-energy
constants are from table 1. The final predictions are evaluated with the
amplitudes shown in table 4, using
$\bar{f}=5.59(1+0.08q^2),\bar{g}=4.77(1+0.08q^2)$. For the evaluation of the
uncertainties in the
rates see text.}

\begin{center}
\vspace{.5cm}
a) $K^+$ decays

\vspace{.2cm}
\begin{tabular}{|c|cccc|}
\hline
 &$\pi^+\pi^- e^+\nu_e$&$\pi^0\pi^0 e^+ \nu_e$&$\pi^+\pi^-\mu^+
\nu_\mu$&$\pi^0\pi^0 \mu^+ \nu_\mu$
\\ \hline
{\small tree}&$1297$& $683$&$155$&$102$\\

{\small one-loop}&$2447$&$1301$&$288$&$189$\\
\hline
{\small final}&{\small input}&
$1625$&$333$&$225$\\
{\small prediction} & &$\pm90$&$\pm15$&$\pm11$\\ \hline

experiment &$3160$& $1700$&$1130$&  \\
\cite{pdg}&$\pm140$&$\pm320$&$\pm730$&\\
\hline
\end{tabular}

\vspace{1cm}
b) $K^0$ decays

\vspace{.2cm}
\begin{tabular}{|c|cc|}
\hline
 &$\pi^0\pi^-e^+ \nu_e$&$\pi^0\pi^- \mu^+ \nu_\mu$ \\ \hline
{\small tree}& $561$&$55$\\

{\small one-loop}&$953$&$94$\\
\hline
{\small final}&
$917$&$88$\\
{\small prediction} &$\pm170$&$\pm22$\\ \hline

experiment & $998$&  \\
\cite{kld}&$\pm39 \pm 43$&\\
\hline
\end{tabular}
\end{center}
\end{table}

 We have assessed an uncertainty due to contributions
from
$F^-_\mtiny{CHPT},R_\mtiny{CHPT}$ in the following manner.
i) We have checked that the results barely change by using the tree level
expression for $R_\mtiny{CHPT}$ instead of its one-loop representation. We
conclude
from this that the uncertainties in $R_\mtiny{CHPT}$ do not matter. ii) The
uncertainty from $F^-_\mtiny{CHPT}$
 is taken into account by adding to $\Delta \Gamma$ in quadrature the
change obtained
by evaluating $F^-_\mtiny{CHPT}$ at $L_3=-3.5+1.1=-2.4$. iii) In $K^0$ decays,
we have also added in quadrature the difference generated by evaluating the
rate with $M_\pi=135$ MeV.

The decay $K^0\rightarrow \pi^0\pi^-e^+\nu_e$  has recently been
measured \cite{kld} with considerably higher statistics than before \cite{pdg}.
We display the result for the rate in the first column of table 5b. The
quoted errors correspond to the errors in the branching
ratio \cite{kld} and do not include the uncertainty  in the total
decay rate  quoted by the PDG \cite{pdg}.
Notice that the  value for $L_3$ determined in \cite{kld} should be multiplied
with $-1$ \cite{makpriv}.

\renewcommand{\theequation}{\arabic{section}.\arabic{equation}}
\renewcommand{\thetable}{\arabic{table}}
\setcounter{equation}{0}
\setcounter{subsection}{0}
\section{Summary and conclusion}
\begin{enumerate}
\item
The matrix elements for $K_{l4}$ decays depend on four form factors $F,G,H$ and
$R$. This article contains the full expressions for these  at
order $E^4$ in CHPT, thus completing already published
calculations \cite{wesszkl4,bijnenskl4,riggen} of $F,G,H$ at this order.

\item
We have estimated higher order terms in the $S$-wave amplitude of the form
factor $F$ by use of a dispersive representation, determining the subtraction
constants in the standard manner \cite{dgl} from CHPT.
 This procedure puts earlier attempts
\cite{riggen} to estimate these corrections on a more firm  basis.

\item
Using the improved $S$-wave amplitude, we have determined $L_1^r,L_2^r$ and
$L_3$ from
$K^+\rightarrow \pi^+\pi^-e^+\nu_e$ decays and $\pi\pi$ threshold data.
 Unitarizing the amplitude
affects the related $SU(2)\times SU(2)$ constant $\bar{l}_1$
in a significant manner. As a result of this,
 the $D$-wave scattering lengths are in better agreement with the
values given by Petersen \cite{nagels} than was the case before \cite{riggen}.
All in all,  a remarkably  good
 agreement with $K_{e4}$ and $\pi\pi$ data is obtained.

\item
 $K_{l4}$ decays may be used to
test the large-$N_C$ prediction $L^r_2=2L^r_1$ \cite{bijnenskl4,riggen}.
 Using the improved
representation of the amplitudes, we have confirmed the earlier \cite{riggen}
finding: The
large-$N_C$ rule works at the one standard deviation level for this
combination of the constants.

\item
The above determination of $L^r_1,L^r_2$ and $L_3$ will presumably be even more
reliable, once high statistics data from kaon facilities like DA$\Phi$NE
\cite{dafne} will become available.

\item
We also predict the slope $\lambda_g$ of the form factor $G$ and total
decay rates, see Eq. (\ref{eslope}) and table 5.

\item
We have made some effort to find out whether any of the $K_{l4}$ decays could
serve to determine some of the other low-energy constants  which
occur in the amplitude. We believe that it will be very difficult to pin down
any of these (in particular $L^r_4$) to better precision than already known,
because the higher order
corrections tend to wash out their effect.

\item
Finally, we would like to recall that the determination of the low-energy
constants from $K_{l4}$ decays or the prediction of the total rates is not the
only issue: these decays are in addition the only known source for a
precise determination of the isoscalar $\pi\pi$ $S$-wave phase shift near
threshold. The possibilities to determine those in future
high statistics experiments are presently under investigation \cite{pipiphase}.

\end{enumerate}

\vspace{3cm}

\noindent
{\bf{Acknowledgements}}\\
It is a pleasure to thank Gerhard Ecker, Marc Knecht and Jan Stern for
enjoyable discussions and Greg Makoff  for communications
concerning  the experiment described in \cite{kld}.

 \newpage
\renewcommand{\theequation}{\Alph{section}.\arabic{equation}}
\appendix

\section{Loop integrals}
\label{loop}
In this appendix we define the loop integrals used in the text.
We consider a loop with two masses, $M$ and $m$.
 All needed functions can be given
in terms of the subtracted scalar integral $\bar{J}(t) = J(t) - J(0)$
evaluated in four dimensions,
 \begin{equation}
J(t) = ~  -i
\int \frac{d^dp}{(2\pi)^d} \frac{1}{((p+k)^2 - M^2)(p^2 - m^2)} \co
\end{equation}
with $t = k^2$.
The functions used in the text are then :
\begin{eqnarray}
\bar{J}(t)&=&-\frac{1}{16\pi^2}\int_0^1 dx~
\log\frac{M^2 - t x(1-x) - \Delta x}{M^2 - \Delta x}
\nonumber\\&=&
\frac{1}{32\pi^2}\left\{
2 + \frac{\Delta}{t}\log\frac{m^2}{M^2} -\frac{\Sigma}{\Delta}
\log\frac{m^2}{M^2} - \frac{\sqrt{\lambda}}{t}
\log\frac{(t+\sqrt{\lambda})^2 -
\Delta^2}{(t-\sqrt{\lambda})^2-\Delta^2}\right\}
{}~,
\nonumber\\
J^r(t) &=& \bar{J}(t) - 2k~,
\nonumber\\
M^r(t) &=& \frac{1}{12t}\left\{ t - 2 \Sigma \right\} \bar{J}(t)
+ \frac{\Delta^2}{3 t^2} \bar{J}(t)
+ \frac{1}{288\pi^2} -\frac{k}{6}
\nonumber\\&&
                               - \frac{1}{96\pi^2 t} \left\{
      \Sigma + 2 \frac{M^2 m^2}{\Delta}
     \log\frac{m^2}{M^2} \right\} ~,
\nonumber\\
L(t)&=& \frac{\Delta^2}{4t} \bar{J}(t)~,
\nonumber\\
K(t)&=&\frac{\Delta}{2t}\bar{J}(t)    ~,
\nonumber\\
\Delta &=& M^2 - m^2~,
\nonumber\\
\Sigma &=& M^2 + m^2  ~,
\nonumber\\
\lambda&=&\lambda(t,M^2,m^2) ~=~ (t+\Delta)^2 - 4tM^2  ~.
\end{eqnarray}
In the text these are used with subscripts,
\begin{equation}
\bar{J}_{ij}(t)  =  \bar{J}(t)~~~\mbox{with}~~~M = M_i , m = M_j~ \co
\end{equation}
and similarly for the other symbols.
The subtraction point
dependent part
is contained in the constant $k$
\begin{equation}
k = \frac{1}{32\pi^2} \frac{M^2 \log \left( \frac{M^2}{\mu^2} \right)
                       - m^2 \log\left(\frac{m^2}{\mu^2}\right)}
   {M^2 - m^2},
\end{equation}
where $\mu$ is the subtraction scale.

\setcounter{equation}{0}
\section{Resonance contribution to the form factors}
\label{rescont}
Below we display the contributions to the form factors $F$ and $G$
from  resonance exchange (spin less than or equal to one, see also
\cite{ko,finke,eck}). We quote them for $K^+ \rightarrow \pi^+ \pi^-l^+\nu_l$.
The others can be derived using isospin relations (\ref{i2}).
 These
contributions have been used both to provide
 a reasonable approximation of the
left-hand cut, and to estimate higher order corrections in $\bar{g}$.

\vspace{.5cm}
{\bf VECTORS}
\begin{enumerate}
\begin{itemize}
\item
$t$-channel
\bea
F^t_V&=& \frac{M_K G_V}{2\sqrt{2}F_\pi^3} \frac{1}{M_V^2-t}\times \nn
& & \left[ F_V(t-u+2s_l)+G_V(t-u-3s_\pi-s_l+M_K^2+8M_\pi^2)\right] \co \nn
G^t_V&=& \frac{M_K G_V}{2\sqrt{2}F_\pi^3} \frac{1}{M_V^2-t}\times \nn
& & \left[F_V(M_K^2+s_l-s_\pi)+G_V(t-u+M_K^2+s_\pi-s_l)\right] \co
\eea
\item
$s_\pi$-channel
\bea
F^{s_\pi}_V&=& -\frac{M_K G_V}{2\sqrt{2}F_\pi^3} \frac{1}{M_V^2-s_\pi}
\left[ F_V-2G_V\right](t-u) \co \nn
G^{s_\pi}_V&=& \frac{M_K G_V}{2\sqrt{2}F_\pi^3} \frac{1}{M_V^2-s_\pi}\times \nn
& & \left[ (F_V-2G_V)(s_l-M_K^2)+(F_V+2G_V)s_\pi\right] \co
\eea
\end{itemize}

\vspace{.5cm}
{\bf SCALARS}
\begin{itemize}
\item
$t$-channel
\bea
F_S^t&=&\frac{\sqrt{2}M_K}{F_\pi^3}\frac{1}{M_S^2-t}\times \nn
& & \left[ c_d^2(M_K^2+M_\pi^2-t)-c_d c_m(M_K^2+M_\pi^2)\right] \co \nn
G_S^t&=&-F_S^t
\eea
\item
$s_\pi$-channel
\bea
F_S^{\spi}&=&-\frac{2\sqrt{2}M_K}{F_\pi^3}\frac{1}{M_S^2-\spi}\times \nn
& & \left[ c_d^2(2M_\pi^2-\spi)-2c_d c_m M_\pi^2\right] \co \nn
G_S^{\spi}&=&0
\eea

\end{itemize}

\end{enumerate}

The values used for the couplings $G_V, F_V$ and $c_d, c_m$ are \cite{eck}
\bea
F_V&=& 154 ~\MeV \co \nn
G_V&=&  69 ~\MeV \co \nn
c_d&=&  32 ~\MeV \co \nn
c_m&=&  42 ~\MeV \co
\eea
while for the masses we used
\bea
M_V&=& 770 ~\MeV \co \nn
M_S&=& 985 ~\MeV.
\eea

\setcounter{equation}{0}
\section{Evaluation of $f_L,v_0$ and $v_1$}
\label{calcv0}

$f_L$ is calculated from
\be \label{fvs}
f_L = \frac{1}{4\pi} \int d\Omega\left\{(F_V^+ +F_S^+) + \frac{\sigma_\pi
PL}{X}
\cos \theta_\pi(G_V^-+G_S^-)\right\}  .
\ee
Only the $t$-channel contributes to $f_L$. The $s$-channel has only
singularities
on the right-hand cut. The quantities in Eq. (\ref{fvs}) are defined as:

\bea
F_{V,S}^+& = & \frac{1}{2}\left(F_{V,S}^t+F_{V,S}^t(t\leftrightarrow u)
\right)\co \nn
G_{V,S}^-& = & \frac{1}{2}\left(G_{V,S}^t-G_{V,S}^t(t\leftrightarrow u)
\right)\co
\eea
analogously to (\ref{i3}), see appendix B for $G_{V,S}^t,F_{V,S}^t$.
To evaluate $v_0$ and $v_1$, we impose that the unitarized amplitude
$f=f_L+\Omega v$  matches the chiral one-loop representation
$f_\mtiny{CHPT}$ at the threshold $s_\pi=4M_\pi^2$. We write
\bea
f_{\CHPT}&=&\frac{M_K}{\sqrt{2} F_\pi}( f^{(0)}_{\CHPT} + f^{(2)}_{\CHPT}
 +O(E^4)) \co \nn
f^{(0)}_{\CHPT}&= 1 & \co
\eea
with obvious notation, and have
\be\label{v0}
f^{(0)}_{\CHPT}+f^{(2)}_{\CHPT}=\frac{\sqrt{2}F_\pi}{M_K}f_L +
 \left(1+\Delta\right)\left(v_0^{(0)} +
v_0^{(2)} +v_1^{(0)} s_\pi +O(E^4) \right)\co
\ee
where
\bearr
\Delta&=& \frac{s_\pi}{\pi}\int_{4M_\pi^2}^{\Lambda^2} \frac{ds}{s}
\frac{\delta^0_0(s)}{s-s_\pi} \co \nn
 \delta^0_0&=&\frac{2s-M_\pi^2}{32\pi
F_\pi^2}\sqrt{1-4M_\pi^2/s}\co
 \eearr
and where $\delta_0^0$ is a quantity of order $E^2$. The quantity $1+\Delta$
is the expansion of $\Omega$ in CHPT to the required order. The $v_i^{(k)}$ are
obtained by equating the threshold expansion of the left-and right-hand side in
(\ref{v0}).

 \newpage
\addcontentsline{toc}{section}{\hspace{1cm}Bibliography}

\newpage

{ \Large\bf{ {Figure captions}}}

\vspace{1cm}
\begin{enumerate}
\item[Fig. 1]
Kinematic variables for $K_{l4}$ decays. The angle $\theta_\pi$ is
defined in $\Sigma_{2\pi}, \theta_l$ in $\Sigma_{l\nu}$ and $\phi$ in
$\Sigma_K$.

\item[Fig. 2]
The partial wave amplitude $\bar{f}(s_\pi,s_l=0)$. The
dashed line shows $\bar{f}$, evaluated according to Eqs.
(\ref{un5},\ref{fvs}), with $L_1^r,L_2^r$ and $L_3$ from (\ref{dl3}).
 The dash-dotted line is evaluated with
$f_L=0$ according to (\ref{un6}), using the same $L_i$'s, whereas the
solid  line displays $f_s = 5.59(1+0.08q^2). $

\item[Fig. 3]
Differential rate $d\Gamma/ds_\pi$ for $K^+\rightarrow
\pi^+\pi^-e^+\nu_e$ decays in arbitrary units. The evaluation is done with
$F_1^-=F_2=F_3=F_4^-=0$, and
$F_1^+=X  \bar{f}(s_\pi,s_l=0),F_4^+=-PL/XF_1^+$.
The input for the dashed (dash-dotted) line is the same as for the dashed
(dash-dotted) line in Fig. 2.
\end{enumerate}
\end{document}